\documentclass[pra, a4paper, twocolumn, 10pt, showpacs]{revtex4}
\usepackage{bbm, amsmath, amssymb, amsthm, bm,textcomp, nicefrac}
\usepackage[T1]{fontenc}
\usepackage[latin9]{inputenc}
\usepackage{graphicx,graphics}
\usepackage{geometry}
\newcommand{\ket}[1]{\left|{#1}\right\rangle}

\newcommand{\ketbrad}[1]{\left|{#1}\rangle\!\langle{#1}\right|}

\newcommand{\be}{\begin{equation}}
\newcommand{\ee}{\end{equation}}
\newcommand{\eea}{\end{eqnarray}}
\newcommand{\bea}{\begin{eqnarray}}

\begin{document}

\title{Quantum simulation of classical thermal states}

\author{W.\ D\"ur$^1$ and M. Van den Nest$^2$}

\affiliation{$^1$ Institut f\"ur Theoretische Physik, Universit\"at
  Innsbruck, Technikerstr. 25, A-6020 Innsbruck,  Austria.\\
  $^2$Max-Planck-Institut f\"ur Quantenoptik, Hans-Kopfermann-Str.~1, D-85748 Garching, Germany}
\date{\today}

\begin{abstract}
We establish a connection between ground states of local quantum Hamiltonians and thermal states of classical spin systems. For any discrete classical statistical mechanical model in any spatial dimension, we find an associated quantum state such that the reduced density operator behaves as the thermal state of the classical system. We show that all these quantum states are unique ground states of a universal 5-body local quantum Hamiltonian acting on a (polynomially enlarged) system of qubits arranged on a 2D lattice. The only free parameters of the quantum Hamiltonian are coupling strengthes of two-body interactions, which allow one to choose the type and dimension of the classical model as well as the interaction strength and temperature.
\end{abstract}


\maketitle


\textit{Introduction.---} Classical statistical mechanical models are widely studied in physics and have found applications in various contexts, ranging from magnetism and neural networks \cite{Me87} to lattice gauge theories describing fundamental interactions in nature \cite{Ko79}. Thereby models with arbitrary underlying geometry and different symmetry are included, e.g. Ising or Potts models \cite{Wu83} that have a global symmetry or abelian discrete lattice gauge theories with local symmetries. Many of these models show rich phase diagrams and are known to be related to mathematical problems with large computational complexity \cite{Soxx}. Their inherent complexity and limited experimental accessability, e.g. due to a spatial dimension larger than three, hinders in some cases our understanding of these systems.

Here we propose a {\em quantum simulator}  for thermal states of {\em all} discrete classical spin systems in arbitrary spatial dimension. We show that there exists a universal, local 2D quantum Hamiltonian whose unique ground state contains the thermal state of all discrete classical spin systems, including systems with long-ranged and many-body interactions in arbitrary geometries. The choice between different classical models and parameters is provided by the coupling strength of two-body interactions in the 5--body quantum Hamiltonian. For all reasonable discrete classical models, the size of the quantum system is only polynomially enlarged. This opens the possibility to study thermal states of classical spin models in arbitrary dimension using a well-controlled 2D quantum system.

The result is inspired by recently established connections  between classical spin systems and quantum systems \cite{Li97,Ve06,Yu10,Va08,De09}, which have been utilized e.g. to identify complete classical spin models such as a 4D lattice gauge theory, i.e. models from which the partition function of all other classical models can be obtained \cite{Va08,De09,Xu10}. Here we prove a similar completeness result, where we however provide a link between all discrete classical spin models and ground states of local {\em quantum} Hamiltonians. The relation is not only at the level of the partition function, but the thermal state of the classical spin system is indeed contained in the quantum system in a physical sense.

To show this relation, we proceed in three steps: (i) We first introduce  a new mapping between classical spin systems and quantum states, where for each discrete classical spin model a (polynomially enlarged) quantum state is defined that contains the thermal state of the classical model in a certain subsystem, i.e. by simply tracing out or ignoring the additional quantum particles. (ii) We then make use of results from measurement-based quantum computation \cite{Ra01,Br09}, which shows how to obtain arbitrary quantum states from a universal state, the so--called 2D cluster state \cite{Ra01b}. We use these insights to write the quantum state corresponding to any classical spin system as a polynomially enlarged deformed 2D cluster state, where (non--unitary) deformations act on individual qubits. (iii) Finally we show that these deformed cluster states can always be written as the ground state of a universal, local 5-body 2D quantum Hamiltonian, where only coupling strengths of two-body interactions appear as free parameters. The latter technique which allows one to (approximately) write all efficiently preparable quantum states as unique ground states of local Hamiltonians might also find applications in other contexts.


\textit{Mappings between classical spin systems and quantum states.---} Consider a classical spin model with $N$ $2$--level systems ${\bm s}=(s_1,s_2, \ldots s_N)$, $s_i \in \{0,1\}$ which contains \emph{all} many-body interactions of Ising type between groups of at most $d$ spins. For notational simplicity we will here restrict attention to $d =3$ but the generalization to higher $d$ is straightforward. The system is described by the Hamiltonian
\be
\label{H}
H({\bm s})= \sum_a J_a(s_a)+ \sum_{ab} J_{ab}(s_a,s_b) + \sum_{abc} J_{abc}(s_a,s_b,s_c).
\ee
The function $J_a$ specifies an arbitrary local energy term (magnetic field) depending on $s_a$ and $J_{ab},J_{abc}$ associate to each spin configuration of a subset of $2$ or $3$ spins an energy that only depends on the parity of the involved spins, e.g. $J_{ab}(s_a,s_b)=(-1)^{s_a \oplus s_b} J_{ab}$ (which we call an Ising-type interaction). The Hamiltonian $H$ contains all possible 1-, 2- and 3-body interactions of this kind. Hence, by setting suitable couplings to zero, Ising-type models on arbitrary lattices with $3$-body interactions can be obtained.

Interestingly, Ising-type Hamiltonians $H$ are in fact sufficient to treat also non-Ising models with arbitrary $k$-body interactions and $q$-level systems i.e. $H$ defines a \emph{complete} classical spin model, as shown in \cite{De09}. Roughly speaking, one notes that $q$--level spins can be encoded into $\lceil \log_2q \rceil$ 2-level spins, and that arbitrary $k$--body interactions of $q$--level spins can be decomposed into (all possible) $m$--body Ising-type interactions with $m \leq k \times \lceil \log_2q \rceil$. The required Ising interaction strengths can be determined by solving a system of linear equations \cite{De09}. As long as $q$ and $k$ are bounded, this type of encoding is efficient as the overhead in additional qubits scales polynomially with $N$.

We now  define a quantum state that ``contains'' the thermal state of the classical system $H$. We proceed in two steps. First we associate a qubit with every classical spin (vertex particles) and with every interaction term $J_{ab}(s_a,s_b),J_{abc}(s_a,s_b,s_c)$ (interaction particles) and we introduce a  so-called $d$-clique state that encodes the interaction pattern (see also \cite{Va08}) of $H$:
\be
\label{phi}
|\varphi\rangle = \sum_{\bm s} \bigotimes_{abc} |s_a \oplus s_b \oplus s_c\rangle \bigotimes_{ab} |s_a \oplus s_b\rangle \bigotimes_a |s_a\rangle.
\ee
Notice that $|\varphi\rangle$ is a stabilizer state. In the second step, we diverge from the approach of \cite{Va08} and add information about interaction  strengths and inverse temperature $\beta = (k_B T)^{-1}$ via (non-unitary) deformations. We define diagonal matrices $\Lambda_a$, $\Lambda_{ab}$, $\Lambda_{abc}$, where e.g. $\Lambda_{ab}= \mbox{ diag}(e^{-\beta J_{ab}/2}, e^{\beta J_{ab}/2})$, and analogous for $\Lambda_a,\Lambda_{abc}$. We use the short-hand notation $\mathbf{ \Lambda} = \bigotimes_{abc}\Lambda_{abc}\bigotimes_{ab}\Lambda_{ab} \bigotimes_{a}\Lambda_a$ for the tensor product of all single--qubit deformations. We consider the  deformed stabilizer state
\be
|\varphi_{\mathbf{\Lambda}}\rangle = \mathbf{\Lambda} |\varphi\rangle/\sqrt{\cal Z}.
\ee
Here the partition function ${\cal Z} = \sum_{\bm s} e^{-\beta H({\bm s})}$ appears as a normalization. What is more, if one traces out all interaction particles, the resulting density matrix of the vertex particles has as diagonal entries the Boltzmann weights of $H$ \cite{noteBoltzmann}, $\rho={\rm Tr}_{ab,abc}\ketbrad{\varphi_{\mathbf{\Lambda}}}$ with
\be
\langle
{\bm s}|\rho|{\bm s}\rangle = e^{-\beta H({\bm s})}/\cal Z.
\ee
Therefore all classical quantities such as spin correlations or energies are contained in $\rho$ and can be read out using diagonal quantum observables.

We remark that the states $|\varphi_{\mathbf{\Lambda}}\rangle$ can be written as ground states of certain $d$--body quantum Hamiltonians. However, these Hamiltonians may be highly non-local and may contain $O(N)$-body interactions terms. In the following we will show how to relate this state to a local \emph{two-dimensional} Hamiltonian. To this aim it is crucial that the deformations in $|\varphi_{\mathbf{\Lambda}}\rangle$ are single-qubit operators, which is ensured by the specific mapping above making use of only Ising type interactions.


\textit{Universal quantum state.---} We now show that, independent of the choice of parameters, any state $|\varphi_{\mathbf{\Lambda}}\rangle$ can be written to arbitrary accuracy as a deformed 2D-cluster state with invertible single--qubit deformation operators. The 2D cluster state $|{\cal C}\rangle$ \cite{Ra01b} is  a universal resource for measurement-based quantum computation (MQC), in the sense that any quantum state can be obtained from a sufficiently large 2D-cluster state by means of single-qubit measurements \cite{Ra01,Br09}. The 2D-cluster state is associated with a 2D rectangular lattice by placing qubits on the vertices, and is defined as
\be
\label{clusterstate} |{\cal C}\rangle = {\cal U} |+\rangle^{\otimes M},\quad {\cal U}=\prod_{(a,b) \in E} U_{ab}
\ee
where $|+\rangle=\frac{1}{\sqrt{2}}(|0\rangle + |1\rangle)$, $U_{ab}= \mbox{ diag} (1,1,1,-1)$ and $E$ are the edges of the 2D lattice. The universality of the 2D cluster state implies that there exists single-qubit states $|\omega_a\rangle$ such that
\be
\left (\bigotimes_{a \in A} \ketbrad{\omega_a} \otimes I \right ) |{\cal C}\rangle = \frac{1}{2^{\frac{|A|}{2}}}\bigotimes_{a \in A} \ket{\omega_a} \otimes |\varphi\rangle,
\ee
i.e. by projecting out all particles in $A$, the stabilizer state $|\varphi\rangle$  can be generated at the remaining particles. Notice that this method is constructive and efficient, i.e. a polynomially enlarged 2D cluster state suffices and only Pauli measurements are involved since $|\varphi\rangle$ is a stabilizer state. In order to relate the resulting quantum state to a local Hamiltonian, we now replace the projection operators $\ketbrad{\omega_a}$ by invertible operators $\Omega_a = (1-\epsilon) \ketbrad{\omega_a} + \epsilon \ketbrad{\omega_a^\perp}$, where a polynomially (in $N$) small $\epsilon$ suffices to guarantee an approximation of the desired state $|\varphi\rangle$ with polynomial accuracy
\cite{noteFT}.
To be precise, the resulting state after applying imperfect projectors $\Omega_a$ at each of the $|A|$ qubits is given by a sum of $2^{|A|}$ terms, corresponding to all possible combinations where at $n$ positions the desired projection $\ketbrad{\omega_a}$ has been applied, while at $|A|-n$ positions the (wrong) projector $\ketbrad{\omega_a^\perp}$ was applied. Every term has a weight $(1-\epsilon)^n\epsilon^{|A|-n}$, while each of the resulting states has the same norm, independent of which projection operators have been applied. The latter point is a special feature of 2D cluster states that also occurs in measurement-based quantum computation, where outcomes of projective measurements in any basis occur with equal probability at any stage of the computation. The reason is that reduced density operator of each qubit is given by the identity due to the entanglement with the remaining (unmeasured) qubits. We can then lower bound the fidelity of the resulting state as compared to the one obtained by the perfect projections by $F \geq (1-\epsilon)^{|A|}$, where we assume that only the term where all projections lead the desired outcome contribute to the fidelity.
Using the notation $\mathbf{\Omega} = \bigotimes_{a \in A} \Omega_a$, it follows that the  deformed stabilizer state $|\varphi_{\mathbf{\Lambda}}\rangle$ can be written to arbitrary accuracy as a deformed 2D cluster state
\be
\label{deformedcluster}
|{\cal C}_{\mathbf{\Omega, \Lambda}}\rangle \equiv \mathbf{\Omega} \otimes \mathbf{\Lambda} |{\cal C}\rangle \approx \bigotimes_{a \in A} \ket{\omega_a} \otimes |\varphi_{\mathbf{\Lambda}}\rangle,
\ee
where we have omitted the normalizations. Notice that there are 3 different groups of qubits: group $A$ are auxiliary qubits, which are deformed by $\mathbf{ \Omega}$ to form  the state $|\varphi_{\mathbf{\Lambda}}\rangle$ on the remaining qubits. The second group $B$ are all interaction qubits corresponding to Ising-type interactions, where the deformation is determined by the interaction strengths $J_{ab}$ and $J_{abc}$ and the inverse temperature $\beta$. Finally, the group of vertex qubits $C$ is acted upon by deformations determined by the local fields $J_a$ and by $\beta$. At the same time, $C$ is the group on which the thermal state of the corresponding classical system in generated.


\textit{Ground states of local quantum Hamiltonians.---} We now show that the deformed cluster state $|{\cal C}_{\mathbf{\Omega, \Lambda}}\rangle$ can always be written as the unique ground state of a universal local 5--body Hamiltonian of a 2D quantum system.

The deformations $\mathbf{\Omega}$ and $\mathbf{\Lambda}$ are treated  in different ways. We first consider the partially deformed state $|{\cal C}_{\mathbf{\Lambda}}\rangle=I \otimes \mathbf{\Lambda}|{\cal C}\rangle$. Note that $\mathbf{\Lambda}$ (which acts on groups $B$ and $C$) is a {\em diagonal} matrix which hence commutes with the phase gates ${\cal U}$ so that $|{\cal C}_{\mathbf{\Lambda}}\rangle={\cal U} [I \otimes \mathbf{\Lambda}]|+\rangle^{\otimes N}$. Remark that for every $i\in A$ the state $|+\rangle_i$ is the unique ground state of $-X_i$ with ground state energy $-1$. Moreover each of the local deformations transforms $|+\rangle_k$ with $k\in B\cup C$ in $e^{-\beta J/2} |0\rangle +  e^{\beta J/2} |1\rangle$. This state is the unique ground state of a single-qubit Hamiltonian $ -X_k - \gamma_k Z$ for some suitable $\gamma_k$ with ground state energy, say, $E_k$. For every $a$ consider the 5-local operator $K_a = X_a \prod Z_{b}$ where the product is over all neigbors $b$ of $a$ in the 2D lattice; the $K_a$ are the standard stabilizer operators of the cluster state. Using that ${\cal U}X_a{\cal U}^\dagger = K_a$ and ${\cal U}Z_a{\cal U}^\dagger = Z_a$, we find that $|{\cal C}_{\mathbf{\Lambda}}\rangle$ is the unique state satisfying
\bea
(I - K_i)|{\cal C}_{\mathbf{\Lambda}}\rangle = 0 \quad (- K_k - \gamma_k Z_k - E_k I)|{\cal C}_{\mathbf{\Lambda}}\rangle  =0
\eea
for every $i\in A$ and $k\in B\cup C$, and is hence the zero-energy ground state of the Hamiltonian containing the sum of all these terms.

We now deal with the deformations $\mathbf{\Omega}$ acting on
group $A$. The  operator $\mathbf{\Omega}$ is in general not
diagonal, so a different treatment is necessary. First, let
$\mathbf{\Omega}^i$ denote the tensor products of all $\Omega_a$
which act on  $i$ and its nearest-neighbors in the 2D lattice.
Since $|{\cal C}_{\mathbf{\Lambda}}\rangle$ is a zero energy
ground state of $I-K_i$, it follows that $|{\cal
C}_{\mathbf{\Omega, \Lambda}}\rangle$ is a zero energy ground
state of $P_i = [\mathbf{\Omega}^i]^{-\dagger} (I-K_i)
[\mathbf{\Omega}^i]^{-1}$ for every $i\in A$ \cite{Ciracprivate}.
This particular choice of transformation guarantees that the
resulting Hamiltonian $P_i$ is hermitian, and remains $5$--body
\cite{noteH}. Notice that here it is important that the inverse of
$\mathbf{\Omega}^i$ exists, which is the reason why we approximate
projection operators by invertible operators $\Omega_b$ in Eq.
(\ref{deformedcluster}). Second, we consider the terms $H_k=- K_k
- \gamma_k Z_k- E_k I$, $k \in B \cup C$ of the Hamiltonian, where
in each term only one particle $n_k \in A$ is affected by a
deformation in group $A$ (see Fig. 1b). Hence $|{\cal
C}_{\mathbf{\Omega, \Lambda}}\rangle$ is  a zero-energy ground
state of $Q_k = [{\Omega}_{n_k}]^{-\dagger} H_k
[{\Omega}_{n_k}]^{-1}$. Putting everything together, we find that
$|{\cal C}_{\mathbf{\Omega, \Lambda}}\rangle$ is the unique ground
state of the 2D 5-body Hamiltonian \bea \label{universalH} {\cal
H} = \sum_{i \in A}  P_i + \sum_{k \in B\cup C} Q_k. \eea Notice
that term $-\gamma_k Z_k$ in $H_k$ is transformed into a two-body
interaction $\gamma_k ([{\Omega}_{n_k}]^{-\dagger}
[{\Omega}_{n_k}]^{-1}) \otimes Z_k$. The strength of this fixed
two-body term is determined by the parameters of the deformation
$\mathbf{\Lambda}$ (and hence the parameters of the corresponding
classical model). By the completeness of the classical model
(\ref{H}), one can thus change between all classical models (which
can correspond to arbitrary geometries and also to local or global
symmetries) by simply adjusting the strength of these two-body
interactions. Notice that in this way we obtain a constructive
method to realize thermal states of arbitrary classical models in
{\em any} spatial dimension as ground states of a 2D local quantum
Hamiltonian. This proves the main result of this paper.

Note that a reduction to a local 4--body Hamiltonian is possible by using a universal graph state corresponding to a 2D hexagonal lattice instead of the 2D cluster state.

\begin{figure}[ht]
  \begin{picture}(210,95)
\put(-22,-90){\includegraphics[width=9cm]{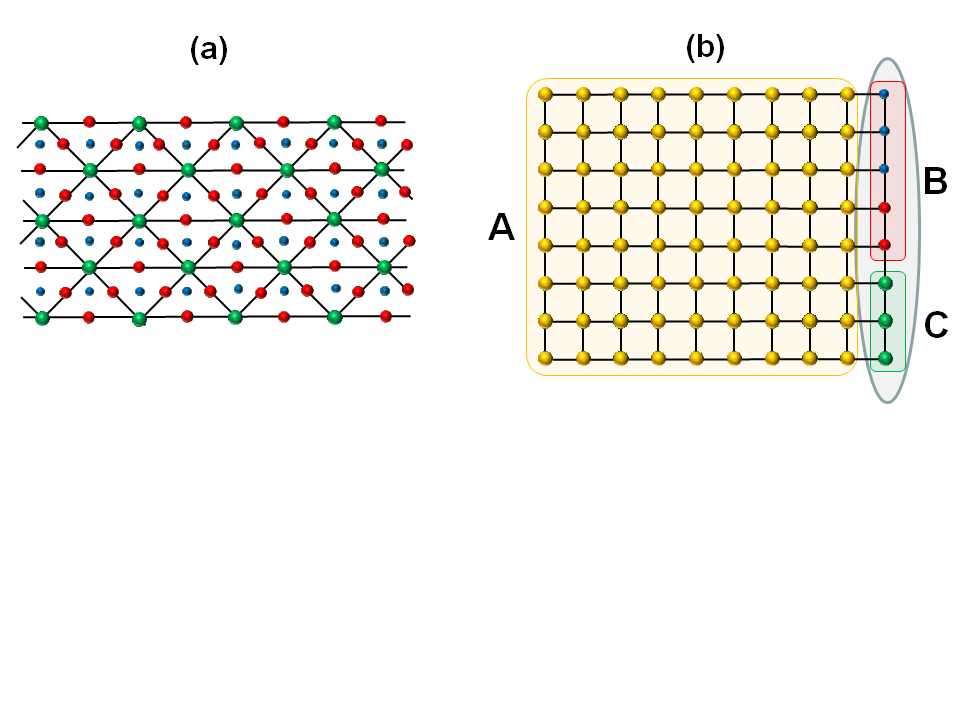}}
\end{picture}
\caption[]{\label{Triangle} (a) Illustration of mapping for a classical spin system on a triangular lattice with nearest neighbor 2-body and 3-body plaquette interactions. Quantum particles are associated with vertices (green, large), nearest neighbor pairwise interactions (red, medium) and 3-body plaquette interactions (blue, small). The classical thermal state is generated at the vertex particles (green, large). (b) 2D Cluster state with groups $A$, $B$, $C$. Deformations on group $A$ allow one to generate the state $|\varphi\rangle$ on system $B \cup C$ (Fig. 1(a)), while deformations on group $B$ and $C$ allow one to chose interaction strengthes and temperature.
}
\end{figure}

{\it Direct constructions.---}  It is straightforward to repeat the above argument starting from Ising-type models on arbitrary lattices rather then the complete model (\ref{H}). This involves a stabilizer state similar to $|\varphi\rangle$ which now encodes the interaction pattern; see Fig \ref{Triangle} for an example. The application of the deformations $\mathbf{\Lambda}$ is analogous, as is the mapping to a deformed cluster state which is then found to be a ground state of a 2D 5-body Hamiltonian as well.

\textit{Discussion and Conclusion.---} We have shown that there exists a universal family of 2D 5-body quantum Hamiltonians Eq. (\ref{universalH}) whose unique ground states include the thermal states of all classical spin models in arbitrary spatial dimension. For all classical spin models with $q$--level spins and $k$--body interactions where $q$ and $k$ are bounded --including edge models, vertex models as well as lattice gauge theories--, the size of the corresponding quantum system is polynomially enlarged, and the classical thermal state is generated at a subsystem of the quantum spins. The parameters and type of the classical model are determined by the strength of fixed two-body interactions in the quantum Hamiltonian.

Apart from this general result, two techniques applied in this proof are of particular relevance and might find applications elsewhere. On the one hand, the new mapping relating classical models with deformed stabilizer states allows one to study the feature of classical models by investigating the (entanglement) properties of the corresponding deformed stabilizer states. On the other hand, the technique of relating a deformed cluster state to a ground state of a local Hamiltonian also allows one to write quantum states as ground states of a local quantum Hamiltonian. It would be interesting to see if the possibility to efficiently prepare these states via MQC is reflected in the properties of the corresponding Hamiltonian, e.g. its gap.

We also point out that our results lead to a quantum simulator for classical thermal states. This is either based on the direct generation of the ground states of the local quantum Hamiltonian Eq. (\ref{universalH}), e.g. by a cooling process or by adiabatic quantum computation, or by the generation of the states $|\varphi_\Lambda\rangle$ by some other means, e.g. using a quantum circuit. For many classical models this yields an efficient quantum algorithm, e.g. for (inhomogenous) classical models with pairwise interaction pattern given by a graph with bounded rank width, e.g. 1D systems or tree graphs where the corresponding quantum state is a tree tensor network state \cite{Va07}, or the the 2D Ising model without fields (see \cite{Yu10} for more examples). Notice however that one cannot expect that an efficient preparation of the corresponding quantum state is possible for all models, as this would imply the solution of NP-hard problems using our method. One can rather expect that in such cases the quantum state preparation becomes inefficient. It would be interesting to study the connection between computational complexity of the classical spin models and the possibility to efficiently generate the corresponding deformed cluster states, in particular the conditions under which an efficient generation of the quantum state is possible, e.g. by using recently developed algorithms to efficiently prepare certain projected entangled pair states \cite{Frank}. The gap in the Hamiltonian Eq. (\ref{universalH}) plays a key role in this context, and will be subject of further study.

We finally remark that a similar result to ours might be obtained making use of connections between classical thermal states and quantum states pointed out in \cite{Ve06}, each of which can be written as a ground state of a (non-local) quantum Hamiltonian. A reduction to 2D two-body Heisenberg Hamiltionian with only local fields as control parameters could be achieved by making use of gadgets constructions of Refs. \cite{Ol08,Sc09}. This method however has the disadvantage that already for models with long-ranged two-body interactions, i.e. where each spin interacts with $O(N)$ other spins, the corresponding quantum Hamiltonian is $N$-body. This leads to an exponential overhead in the number of auxiliary particles in the gadget construction. In addition, the gadget constructions put stringent requirements on the necessary accuracy of the control parameters of the Hamiltonians \cite{Ol08,Sc09}.
Notice that gadget constructions can also be applied as a final step in our construction, reducing the universal quantum Hamiltonian Eq. (\ref{universalH}) to an effective two-body Hamiltonian. In addition, the two-body interactions controlling the type and parameters of the classical model can be reduced to a local single-particle magnetic field together with fixed two-body interactions, i.e. the change between different classical models is achieved by varying local fields only.

\textit{Acknowledgements.---} We thank H.J. Briegel, G. De las Cuevas, M.A. Martin-Delgado and B. Kraus for discussions. This work was supported by the FWF and the European Union (QICS, SCALA, NAMEQUAM) and the excellence cluster MAP.


\begin{thebibliography}{99}
\bibitem{noteBoltzmann}
There are also non-zero off-diagonal elements, which can however be easily avoided by adding another copy of vertex qubits in the mapping Eq. \ref{phi}, $\bigotimes_a|s_a\rangle$
\bibitem{noteFT}
Alternatively, one may make use or results from fault tolerant quantum computation, where states can be generated in an encoded form with a fixed error $\epsilon$ in single-qubit measurements. It is however unclear whether the required deformations at the logical level can be implemented by single-qubit operations, which seems necessary to guarantee locality of the resulting Hamiltonian.
\bibitem{noteH}
One may also use the transformation $(I-K_i)\to \mathbf{\Omega} (I-K_i) \mathbf{\Omega}^{-1} = \tilde H_i$ and consider $H_i=\tilde H_i^\dagger \tilde H_i$ to obtain a valid hermitian Hamiltonian.
\end{thebibliography}
\end{document}